\documentclass[amsmath,amssymb,aps,prb,reprint,showpacs,twocolumn,floatfix]{revtex4-1}
\usepackage{graphicx} 
\usepackage{dcolumn}  
\usepackage{times}
\usepackage{hyperref}

\begin{document}

\title{
Structural and electronic changes of pentacene induced by potassium doping
} 

\author{A. Guijarro}
\affiliation{Departamento de Qu\'{\i}mica Org\'anica and
Instituto Universitario de S\'{\i}ntesis Org\'anica, Universidad
de Alicante, San Vicente del Raspeig, 03690 Alicante, Spain.}


\author{J.A. Verg\'es}
\affiliation{Departamento de Teor\'{\i}a y Simulaci\'on de Materiales,
Instituto de Ciencia de Materiales de Madrid (CSIC), Cantoblanco, 28049 Madrid,
Spain.}
\email{jav@icmm.csic.es}

\date{\today}

\begin{abstract}
Potassium is introduced into the crystalline herringbone structure of pentacene searching
for a compound showing metallic electronic transport properties and, hopefully,
superconductivity at small enough temperatures.
Several possible structures for stoichiometric
KPentacene (1:1), K$_2$Pentacene (2:1) and K$_3$Pentacene (3:1) compounds are theoretically investigated.
Detailed densities of states for all of them are presented.
As a more prominent result, a new monoclinic structure has been stabilized
for the potassium richer material that could correspond to the recently synthesized
superconducting phase of K$_3$Pentacene.
Although energetically unfavorable, it is the only metallic candidate found to date.
\end{abstract}

\pacs{71.28.+d, 36.40.Cg, 71.10.Fd, 71.27.+a}

\keywords{Hydrogenation, Polycyclic Aromatic Hydrocarbons, biradical, paramagnetism,
Pariser-Parr-Pople model}

\maketitle

\section{Introduction}

Doping of molecular crystals formed by certain polycyclic aromatic hydrocarbon (PAH)
molecules like picene has originated a new class of superconducting
materials of very promising characteristics. The pioneering work of Mitsuhashi
{\em et al.}\cite{ mitsu2010} showed superconductivity of potassium-intercalated
picene at 7K or 18K depending on sample processing.
Later, other PAH crystals showing a similar laminar structure in which planes are
formed by molecules stacking in a herringbone structure have also shown superconducting
properties according to the temperature behavior of magnetic susceptibility.
Until now, there is only one report of direct measurement of zero resistivity
in this area\cite{ZeroResistivity}.
The review by Kubozono {\em et al.} provides a good starting point to learn about this
emerging field\cite{review}. 

Although parallel to the experimental work there have been many theoretical
calculations of the electronic structure of potassium-intercalated
herringbone structures,
a definite consensus on the precise position of dopant potassium atoms
has not been reached. Experimental samples are not crystalline enough to
allow a precise structural determination using X-ray diffraction.
The lack of this information is particularly relevant when comparing the
doping of picene crystals with the doping of pentacene crystalline samples.
Both elongate molecules are formed
by five benzene rings, picene showing armchair sides that contrast with the
zigzag sides of pentacene. Both are stacked in a herringbone layered structure in the
pristine crystal. Potassium enters within the PAH herringbone structure (intralayer doping)
where there is space enough for three dopant atoms.  
This was the consensus immediately after the discovery of superconducting
properties of K-doped picene\cite{mitsu2010,kos2009,coroneno2011,pedro2011,
giovannetti2011,kosugi2011,kim2011}.

On the other hand, it was widely assumed that doping pentacene with alkaline metals
only improves conductivity for less than one dopant per pentacene molecule.
Potassium, in particular, would occupy interlayer positions donating one electron to
pentacene semiconducting bands. In this way, n-type semiconducting behavior could be
understood\cite{minakata1993}. Some years later, experiments carried out by
Mori and Ikehata on potassium doped pentacene showed an unusual kind of magnetic transition
at low temperatures\cite{mori1997}.
Potassium uptake was near to the 3:1 ratio, but location of potassium atoms was not further
investigated\cite{mori-red}. Quite a few years later,
Hansson {\em et al.} conducted the first numerical research on the
atomic and electronic structures of potassium-doped pentacene\cite{pentaceno-hansson}.
Again small doping
(KPentacene$_2$) would correspond to K interlayer positions but larger doping (KPentacene)
would completely modify the herringbone structure of pristine pentacene:
K would penetrate within the PAH region mutating the layer structure of the
original crystal. The electronic band structures calculated for both compounds show
relatively broad bands that point to a conventional metallic behavior. 
Unfortunately, subsequent experimental work contradicted this explanation:
potassium-intercalated pentacene shows metallic behavior in a broad range of K concentrations
(less than one atom per pentacene molecule) but not for KPentacene\cite{pentaceno-Mott-x1}.
Again, the authors of this work supposed that potassium atoms occupy positions in between the molecular
herringbone layers of pentacene. When interlayer doping is assumed,
there is only place for one K atom per organic molecule because
larger amounts of potassium imply too small distances among dopants.
Actually, the impossibility of having doubly negatively charged pentacene ions
was remarked in Section IV. of Ref.\onlinecite{pentaceno-Mott-x1}.
However,
subsequent experimental works did not recover a Mott-Hubbard transition as a function
of potassium doping but a constant insulator ground state.
Bussolotti {\em et al.} used angle-resolved ultraviolet photoemission spectroscopy to reach
this conclusion by doping of single crystalline pentacene thin films grown on
Cu(110)\cite{pentaceno-Mott-siempre}.
Roth and Knupfer presented electron energy-loss spectroscopy
results for potassium doped samples of tetracene and pentacene that were characterized
by a finite energy gap in the excitation spectra, i.e., none of the samples became
metallic\cite{K2pentacene-mott}. In the last work,
a maximum K doping of two atoms of potassium per pentacene molecule was achieved.

Potassium intercalation of pentacene remains unclear until the present moment.
Two papers published in 2016 present two quite contradictory scenarios.
Phan {\em et al.}\cite{pentaceno-dimeros} report on thin films with the 1:1 stoichiometry.
Napthalene, anthracene, tetracene and
pentacene are potassium intercalated and all of them show insulating character although
the underlying mechanisms are different for short versus long PAHs. On the contrary,
Nakagawa {\em et al.}\cite{nakagawa2016} analyze the properties of a wide range of
K intercalated pentacene samples and observe superconductivity for a large
3:1 stoichiometry. The crystalline structure of the superconducting phase is no longer
triclinic but monoclinic.

Definitely, potassium intercalation of pentacene calls for some clarification. Most
probably, important differences must be associated to structural differences of samples
that are synthesized in different ways. Therefore, theoretical band structure calculations based on
well defined crystalline structures can help to understand different possible scenarios.
Moreover, although the prediction capabilities of Density Functional Theory (DFT) are not
absolute, the possibility of comparing several structures analyzed with exactly the same approach
provides important advantages. 
In our work, we have systematically investigated all the suggested crystalline structures of
potassium intercalated pentacene emphasizing low doping level (just one
K atom per pentacene molecule) and the greatest plausible doping level (tree
K atoms per pentacene molecule).
The possible appearance of magnetic instabilities that can drive the compound to an insulating phase
have been taken into account in our investigation. In this way,
a set of well converged crystalline phases that bear all the structural motives
described so far in the literature has been produced and sorted by their final cell energies.
They have been collected as {\em cif} files in the Supplemental Material\cite{cifs}. 
In particular, an outstanding new phase has been obtained for K$_3$Pentacene that shows
a monoclinic cell as found in Ref. \onlinecite{nakagawa2016} but structurally
quite different from the model proposed in Fig. 6 of this reference.
Our band structure calculation predicts metallic behavior for the new crystalline structure.

\begin{figure}
\includegraphics[width=\columnwidth]{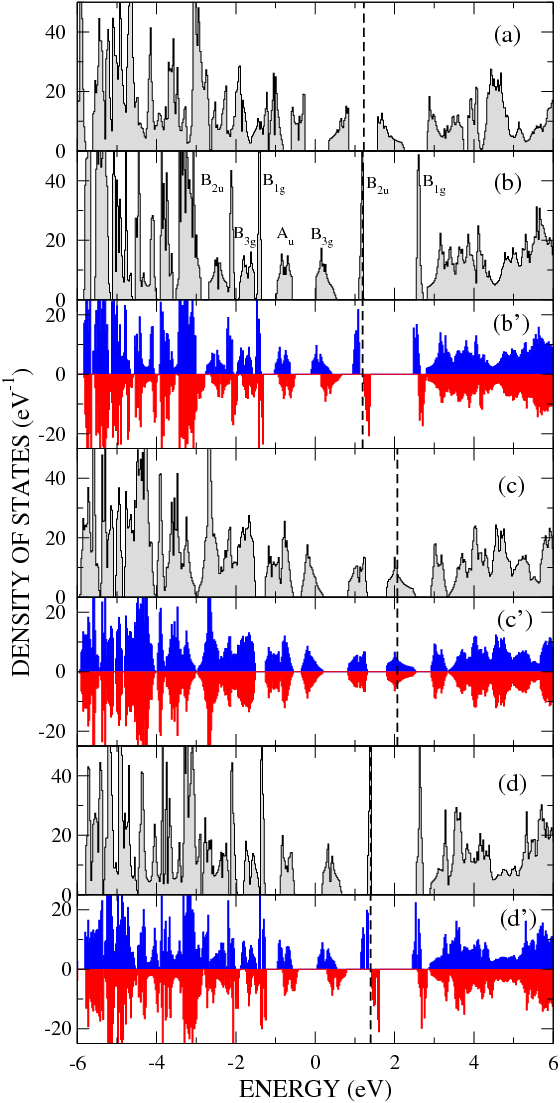}
\caption{(Color online) From top to bottom, (a) density of states (DOS) of
pristine pentacene, (b) DOS for monopotassium pentacene (herringbone structure,
intralayer doping),
(b') same but allowing spin polarization (majority spin polarization positive and
blue colored, minority spin polarization negative and red colored),
(c) DOS for monopotassium pentacene for potassium atoms located between pentacene
layers (interlayer doping) and (c') same but allowing spin polarization, and finally,
(d) DOS for monopotassium pentacene doubling the unit cell to allow the formation of
pentacene dimers bound by two potassium atoms and (d') spin polarized result.
The second panel (b) shows our assignment of some peaks of the DOS to Molecular
Orbitals of pentacene isolated molecule. Fermi levels are indicated by black dashed lines.
}
\label{Kpentaceno}
\end{figure}

\begin{figure}
$\begin{array}{c}
\includegraphics[width=\columnwidth]{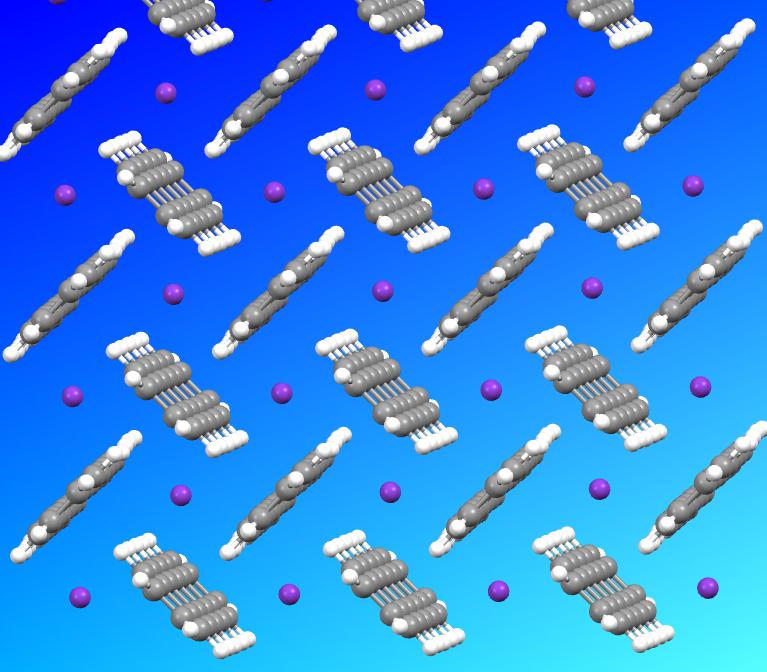}\\
\includegraphics[width=\columnwidth]{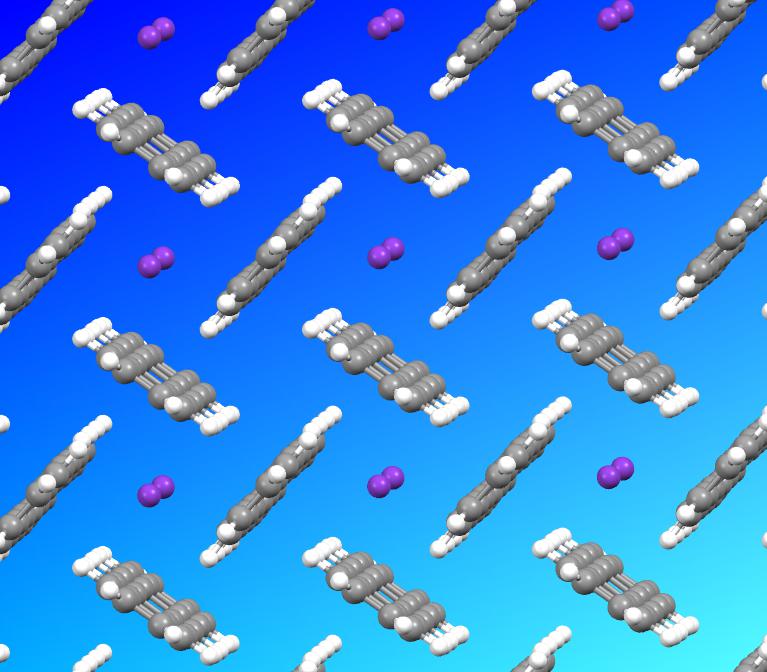}\\
\end{array}$
\caption{(Color online) Herringbone structure of KPentacene (top) compared to the
converged crystalline structure obtained starting from K$_2$Pentacene$_2$ dimers (bottom).
}
\label{estructuras-1}
\end{figure}

The rest of the paper is organized as follows. Section II is devoted to
give some details of the methods and procedures used in this work.
Section III presents our main computational
results together with some discussion of them.
The work ends with a few final concluding remarks (Section IV).

\section{Computational procedures}

Theoretical calculations for model systems of interest have
been performed using {\em ab initio} van der Waals-Density Functional
Theory (vdW-DFT) as proposed in 2004 by Dion {\em et al.}\cite{vdW-DF}.
Taking advantage of an algorithm introduced by Rom\'an-Per\'ez and Soler
the whole calculation can be done in reciprocal space\cite{vdW-DF-num}.
Actually, we have used a later development of this approximation\cite{vdW-DF2}
that was coded by J. Klime\v s within the VASP program\cite{vdW-DF-sol,
vasp1,vasp2,vasp3}.
Wavefunctions have been expanded in a plane-wave basis set
up to a cutoff of $700$\,eV and sampled on a $\Gamma$ centered
Monkhorst-Pack grid automatically generated taking into account the sizes of
the reciprocal lattice vectors (about fifty points during geometry optimizations and
about eight-hundred to get precise densities of states using the tetrahedra method).
Core electrons have been treated within the projector augmented method\cite{paw1,
paw2}. Semi-core 3s and 3p potassium states have been considered valence
states, i.e., they have been fully self-consistently included.
The relaxation of the electronic degrees of freedom has been stopped when both
the total energy and the band structure energy variations
between two steps are  smaller than $10^{-4}$ eV. Ionic relaxation has been
continued as long as any force were larger than 0.01 eV/\AA.

\begin{figure}
\includegraphics[width=\columnwidth]{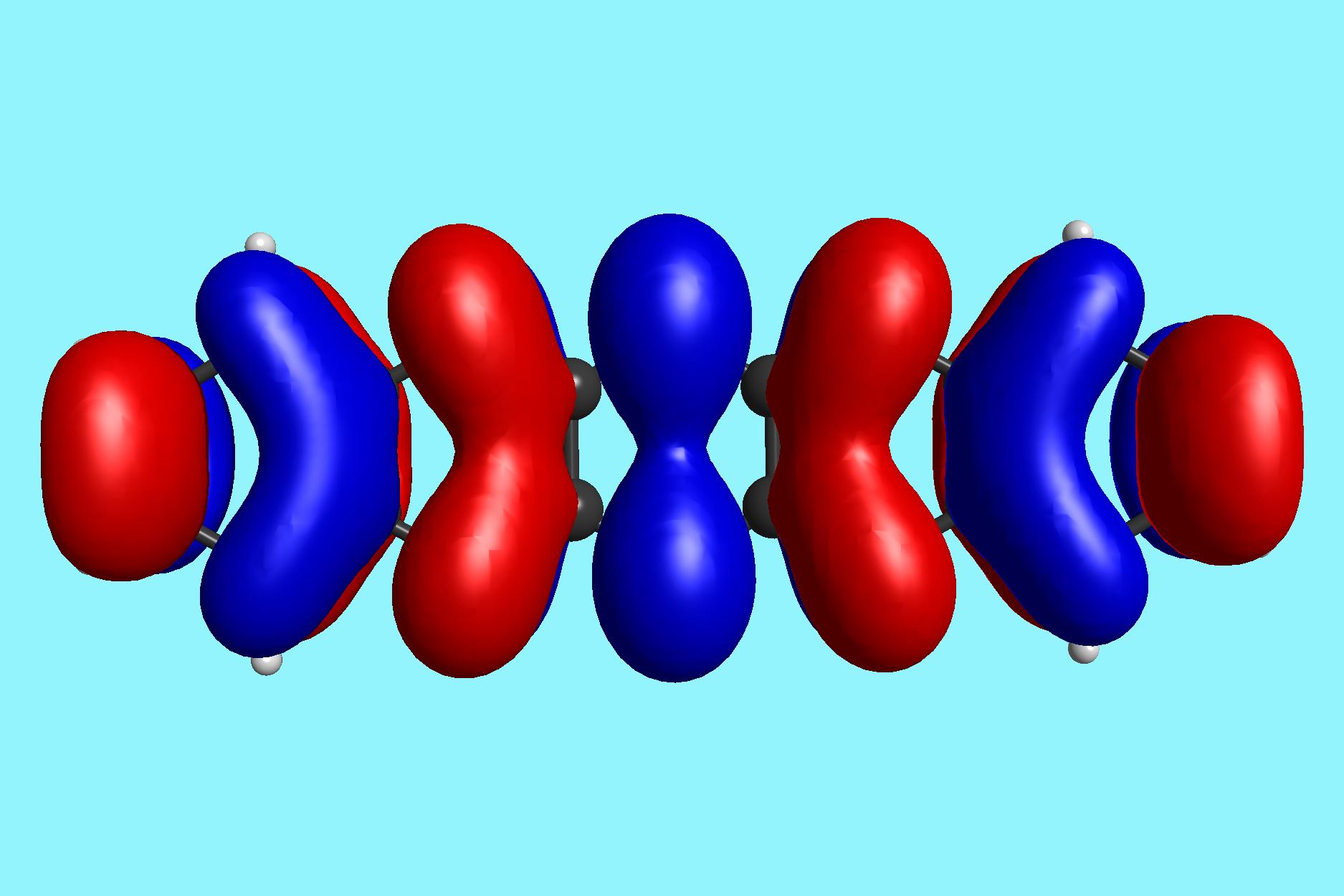}
\caption{(Color online) Picture of LUMO of the pentacene molecule (B$_{\rm 2u}$
symmetry).
}
\label{lumo}
\end{figure}

\begin{figure}
\includegraphics[width=\columnwidth]{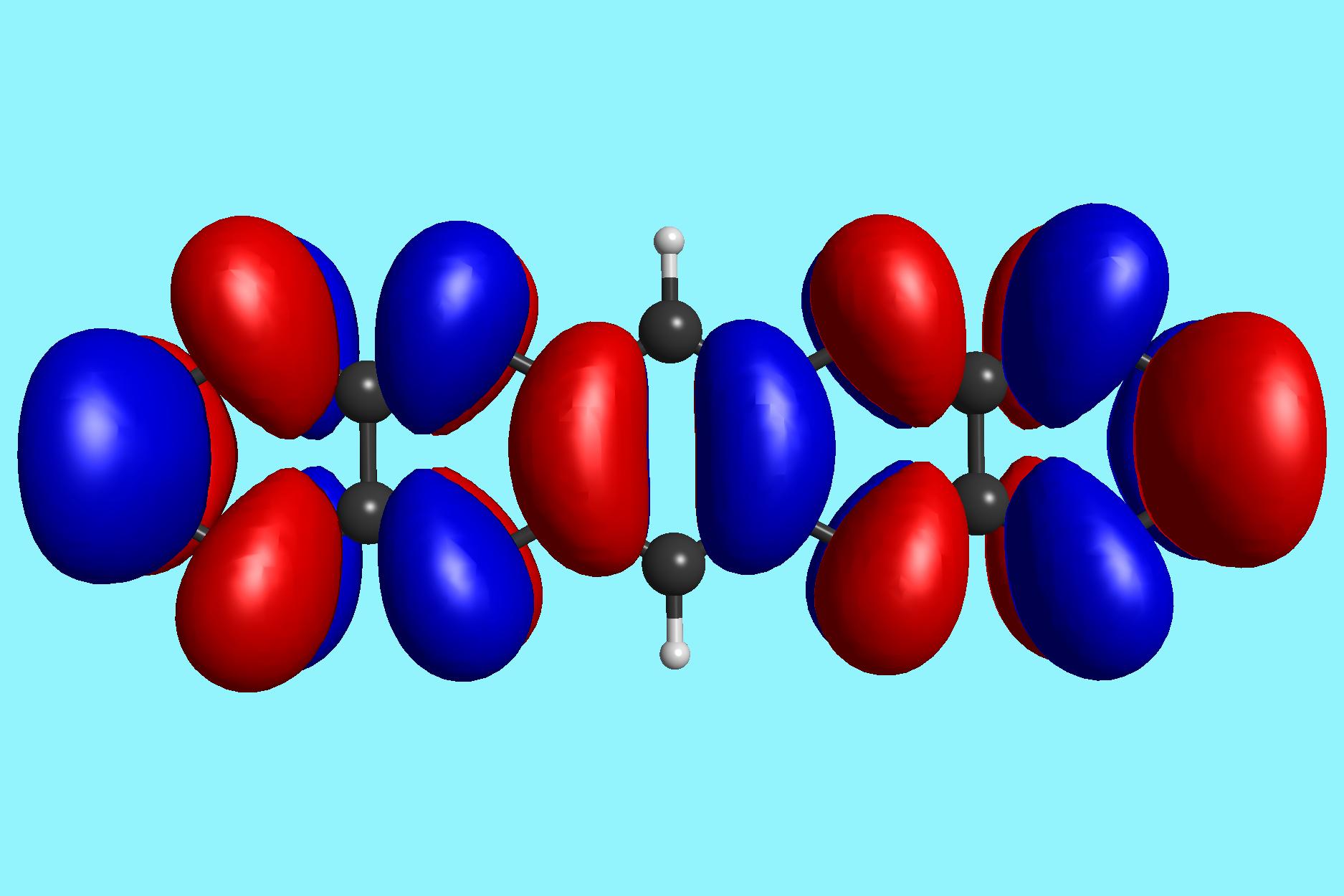}
\caption{(Color online) Picture of LUMO+1 of the pentacene molecule (B$_{\rm 1g}$
symmetry).
}
\label{lumo+1}
\end{figure}

Our whole procedure gives very satisfactory results when it is confronted
with two realistic tests.
Firstly,
polymorphism in pristine pentacene has been studied by calculating the two polymorphs described
for bulk crystalline pentacene\cite{albert}.
Calculations starting from pentacene-HT and pentacene-LT experimental structures afford
the corresponding minima with interlayer spacing of $d(001)$ = 14.4 and 14.1 \AA, respectively.
Optimized structures closely resemble the experimental ones\cite{pentaceno}.
Meanwhile, energy difference between both structures is as small as 3 meV (see Table I).

Secondly,
the structure of the KC$_8$ intercalation compound has been simulated\cite{KC8}.
In this case the behavior of potassium as an intercalation element between carbon
aromatic structures is checked. Theory gives an interplanar distance of 5.48 \AA ~that
compares satisfactorily with the experimental value of 5.35 \AA.
A more accurate comparison can be made using the corresponding {\em cif} files.

We have used VASP Data Viewer to represent iso-surfaces of the electronic density in crystals and
wxMacMolPlt to visualize molecular orbitals of an isolated pentacene molecule\cite{macmolplt}.

\section{Results and discussion}

\begin{table}
\caption{Main properties of the studied K doped pentacene structures.
Volume, spin polarization and total energy are given per cell.
The corresponding {\em cif} files are given as Supplemental Material
(\onlinecite{cifs})}
\begin{ruledtabular}
\begin{tabular}{c|cccc}
Compound & Structure & Volume (\AA$^3$) & Pol. (e) & Energy (eV) \\
\hline
Potassium     &  bcc         & 68.79  &  0  &     1.045 \\
\hline
Pentacene     &  LT          & 688.67 &  0  &  -435.542 \\
              &  HT          & 688.07 &  0  &  -435.539 \\
\hline
              &  intralayer  & 801.57 &  0      & -435.416 \\
              &  intralayer  & 801.57 &  2.007  & -435.505 \\
KPentacene    &  interlayer  & 728.89 &  0      & -435.169 \\
              &  interlayer  & 728.89 &  0.032  & -435.169 \\
              &  dimerized   & 791.20 &  0      & -435.634 \\
              &  dimerized   & 791.20 &  2.006  & -435.718 \\
\hline
K$_{2}$Pentacene  &  herringbone & 816.26  &  0  &  -436.449 \\
\hline
                  &  herringbone & 831.23  &  0      & -434.822 \\
K$_{3}$Pentacene  &  herringbone & 831.23  &  2.007  & -434.874 \\
                  &  monoclinic  & 862.44  &  0      & -434.054 \\
                  &  monoclinic  & 862.44  &  1.060  & -434.053 \\
\end{tabular}
\end{ruledtabular}
\end{table}

\subsection{KPentacene compound}

It is usually believed that when tiny amounts of potassium are introduced
into a pure crystalline sample of pentacene, the original band structure
is not modified but simply populated by the outer 4s electrons of potassium.  
The upper (a) panel of Fig. \ref{Kpentaceno} shows the density of states of pristine pentacene
at the optimized theoretical structure that practically coincides with
the experimentally determined structure at low temperature\cite{pentaceno-fase-LT}.
Two not so narrow bands define the semiconducting gap. As long as this scenario
is valid, the actual positions of dopants are irrelevant since the electric
potential due to potassium cations is not taken into account. In this case, metallic
electronic transport would be possible as long as many-body effects in the form of
a large Hubbard $U$ were absent. Results given in Refs. 
\onlinecite{minakata1993}, \onlinecite{mori1997} and
\onlinecite{pentaceno-Mott-x1} could be understood in this context.
On the other hand, additional experimental
results point to a prompt metal-insulator transition at the first stages
of doping\cite{pentaceno-Mott-siempre, K2pentacene-mott, pentaceno-dimeros}.
A precise description of pentacene compounds with less than
one potassium atom per unit cell
could be theoretically done using crystal supercells of pentacene crystal.
We have not aimed at that goal in this work.
Therefore, we start calculations for a 1:1 K:Pentacene stoichiometry.

If the layers of pentacene showing the herringbone pattern were compact
enough, potassium atoms would occupy interlayer positions. In this way,
typical K-K distances are somewhat larger than 4.5 \AA ~for a KPentacene
compound, that is, quite similar
to the first-neighbor distance in pure crystalline potassium. Consequently,
further doping would be either impossible or giving place to interlayer metallic
potassium (two or more K planes). All these inconveniences are avoided if dopant
atoms penetrate into the herringbone layer producing an intralayer compound (see
upper panel of Fig. \ref{estructuras-1}).
This is certainly the case for picene since doping as high as 3:1 has been
reached. Theoretical estimation of free energies for both structural possibilities
strongly point to intralayer doping. Numerical results compiled in Table I show
that intralayer doping is about 0.7 eV per cell more stable than interlayer
doping. Stabilization is still larger when spin polarized solutions are considered.
Final structures are given as {\em cif} files in the Supplemental Material\cite{cifs}.
Later we will discuss how spin polarization appears and how it strongly suggests
isolating behavior for this compound.

We have also considered the formation of pentacene dimers with two potassium atoms between
hydrocarbon molecules as suggested by Ref. \onlinecite{pentaceno-dimeros}. 
To this end we start forming an isolated K$_2$Pentacene$_2$ dimer that is later used
as the unit of a herringbone structure\cite{tumbados}.
After a long and cumbersome geometry optimization, a conventional herringbone structure is
recovered (see lower panel of Fig. \ref{estructuras-1}). Nevertheless, this lattice shows one important
difference when compared with the previous one since now potassium atoms are grouped in pairs.
Actually, it is still possible to look at this structure as a herringbone array of
Pentacene-K$_2$-Pentacene complexes.
Corresponding cell energies are given in Table I. They indicate that in spite of the reduced
homogeneity this crystalline structure provides the absolute minimum at
a 1:1 stoichiometry by $\approx 0.2$ eV per cell.
This dimerized KPentacene structure could be used as a starting model for a
K$_{0.5}$Pentacene compound (only one K atom per cell) substituting K pairs by
isolated atoms or also for a K$_{1.5}$Pentacene compound if three instead of two
K atoms occupy half of the channels of the herringbone structure.
In any case, going to K richer compounds is not possible since K groups
formed by four or more atoms are not stable.

Searching for additional possible crystalline structures of KPentacene, we have also employed
the structures obtained by Hansson {\em et al.} as unbiased seeds\cite{pentaceno-hansson}.
The resulting converged structures does not differ very much from the initial ones.
However, the calculated total energies are well above
the values of the structures pictured in Fig. \ref{estructuras-1} and, consequently,
they have not been further analyzed. 

\begin{figure}
$\begin{array}{c}
\includegraphics[width=\columnwidth]{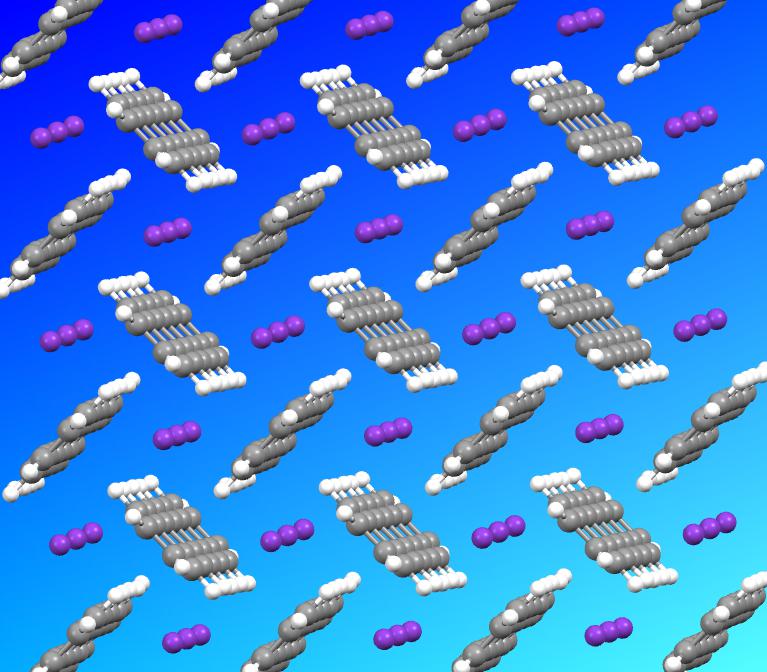}\\
\includegraphics[width=\columnwidth]{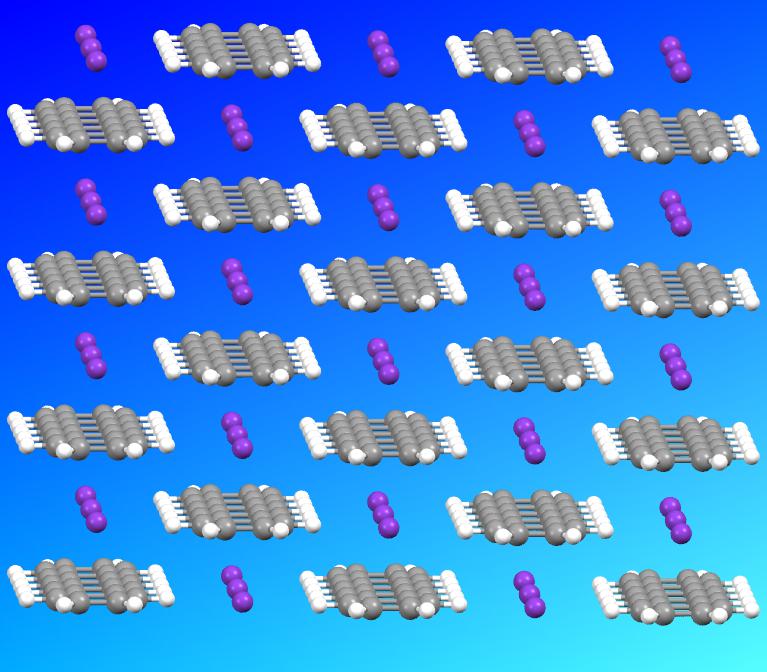}\\
\end{array}$
\caption{(Color online) Herringbone structure of K$_3$Pentacene (top) compared to the new monoclinic
one (bottom).
}
\label{estructuras-3}
\end{figure}

In order to understand electronic transport and magnetic properties of KPentacene, we
have calculated detailed densities of states (DOS) for the relevant structures.
Fig. \ref{Kpentaceno} shows how the change of the bands after doping strongly depend on the precise position
of K dopants. Upper panel shows the DOS of pristine pentacene. Two not so narrow bands define
the semiconducting gap. The second panel gives the DOS for the more stable intralayer structure.
The important narrowing of several bands is quite noticeable. It is due to the symmetry of the corresponding
pentacene molecular orbitals. After structural optimization, pentacene molecules show parallel
long axis and almost perpendicular molecule planes. Molecular orbitals are antisymmetric relative
to the molecule plane since they are linear combinations of $\pi$-orbitals. Therefore, they only overlap
with molecular orbitals of neighboring molecules that are also antisymmetric relative to the symmetry
plane that is perpendicular to the molecule and contains its long axis.
This is not the case for Lowest Unoccupied Molecular Orbital (LUMO)
and LUMO+1 that belong, respectively, to B$_{\rm 2u}$ and B$_{\rm 1g}$
representations of the molecule point symmetry group D$_{\rm 2h}$. Pictures of them are given in
Figs. \ref{lumo} and \ref{lumo+1}.
This property allows an easy assignment of the most relevant bands to specific molecular states.
Since Fermi level divides the B$_{\rm 2u}$ band into occupied and unoccupied parts one could naively think
that the compound would show a metallic behavior. Nothing is further from reality, because electron interaction
leads to spin polarized bands that are completely occupied and completely empty, respectively.
Although our calculation
shows a ferromagnetic solution, i.e., spins of the two doping electrons are parallel, we think that different
spin configurations could compete (for example, anti-parallel spins within the cell). Nevertheless, the
relevant point is that correlation opens a gap in the DOS. In conclusion, structural changes following doping
invalidate a naive picture of electrons populating a relatively broad band (upper panel of
Fig. \ref{Kpentaceno} corresponding to pristine pentacene).

\begin{figure}
\includegraphics[width=\columnwidth]{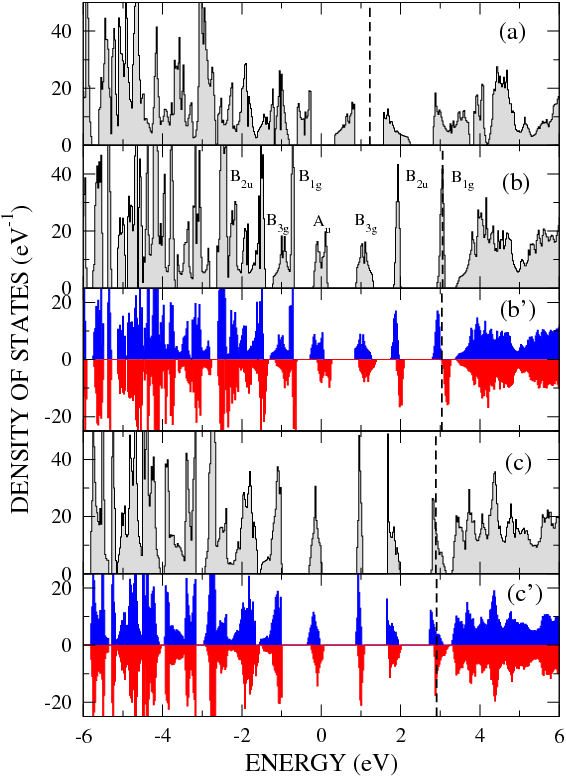}
\caption{(Color online) From top to bottom, (a) density of states (DOS) of
pristine pentacene, (b) DOS for tripotassium pentacene (herringbone structure),
(b') same but allowing spin polarization (majority spin polarization positive and
blue colored, minority spin polarization negative and red colored),
(c) DOS for tripotassium pentacene in the
new monoclinic structure and, (c') same but allowing spin polarization. 
The assignment of some peaks to Molecular Orbitals of pentacene molecules follows the analysis done
for Fig. \ref{Kpentaceno}. Black dashed lines indicate Fermi level positions.
}
\label{K3pentaceno}
\end{figure}

This scenario could be completely different if interlayer doping were possible.
As shown in fourth (c) and fifth (c') panel of Fig. \ref{Kpentaceno},
DOS approximately resembles the undoped bands in this case. The possibility of spin polarization
is marginal for this structural model (see Table I). The position of Fermi level points to a conducting
behavior. Although this structure is energetically unlikely, perhaps the detailed way in which samples
are grown matters and could eventually explain metallic
behavior\cite{minakata1993,mori1997,pentaceno-Mott-x1}.
In any case, as said previously, interlayer doping is limited to one electron (K atom) per pentacene molecule.
Further dopants either form metallic K layers or go into the pentacene herringbone layer.

\begin{figure}
\includegraphics[width=0.9\columnwidth]{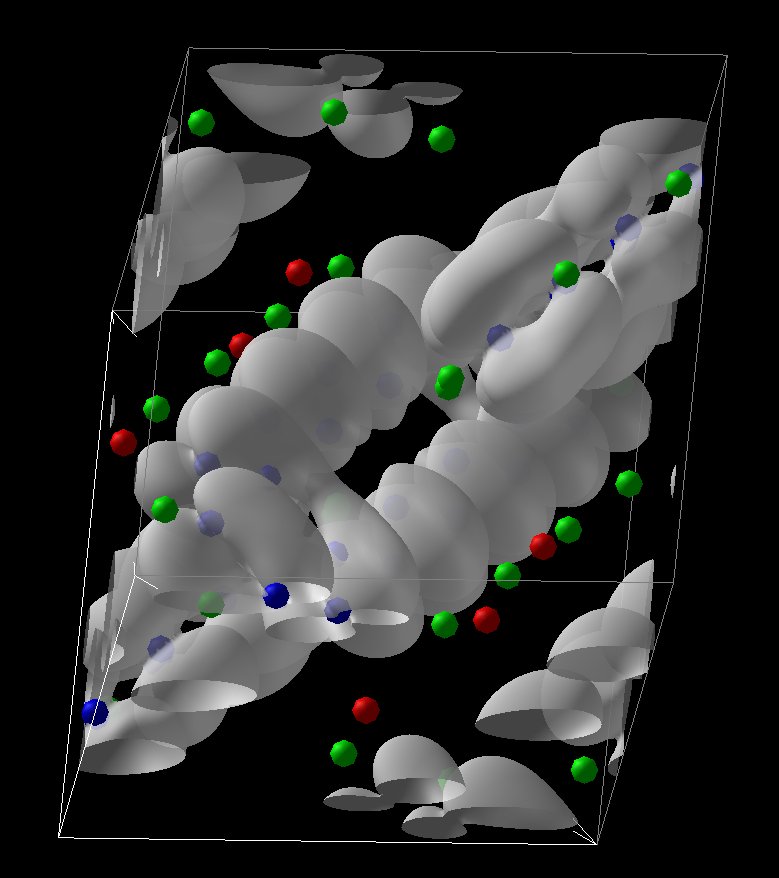}
\caption{(Color online) Iso-surface corresponding to the charge density of
K$_3$Pentacene electrons in bands 127 and 128 (B$_{\rm 2u}$), i.e.,
the first four K electrons donated to pentacene
(iso-surface value is 6).
}
\label{127-128}
\end{figure}

Bottom (d) and (d') panels of Fig. \ref{Kpentaceno} give the density of states calculated for the
dimerized structure. The second one corresponds to the spin polarized solution involving
$\approx 2$ more electrons with up spin than with down spin in every cell. Energy gain is modest,
$\approx 0.08 eV$ per cell. Both figures are very similar to the values obtained for the
more homogeneous solution shown in (b) and (b') panels of Fig. \ref{Kpentaceno}. In fact,
the whole DOS is very similar. This is not a surprise since in both cases LUMO and LUMO+1
states of pentacene describe the relevant bands, being the influence of the precise
position of K cations not so important. Nevertheless, magnetic DOS shows additional small
splittings in the dimerized case. Let us comment here, that the insulating character of our
solution does not come for some kind of Peirls deformation but for large correlation
effects associated to the very narrow half-filled band. This makes a difference with arguments
found in Ref. \onlinecite{pentaceno-dimeros}.  

\subsection{K$_2$Pentacene compound}

It is easy to stabilize pairs of K atoms within the herringbone structure of pentacene. It looks
like the lower panel of Fig. \ref{estructuras-1} once all channels are filled with pairs of
dopants. Complete structural data are given as Supplemental Material\cite{cifs}.
Main characteristics are given in Table I.
The corresponding DOS looks like panel (b) of Fig. \ref{Kpentaceno}
but now the Fermi level is located between B$_{\rm 2u}$ and B$_{\rm 1g}$ peaks. 
This stoichiometry provides maximum lattice stability but not metallic properties. 
Unfortunately, this fact considerably reduces its potential use and
we have not continued its study.

\begin{table}
\caption{Lattice parameters for pentacene (PA) and K-doped pentacene. Spatial
groups are indicated for each structure.}
\begin{ruledtabular}
\begin{tabular}{c|cccc}
  & PA ($P_{-1}$) & KPA ($P_{-1}$) & K$_3$PA ($P_{-1}$) & K$_3$PA ($C_{2/m}$) \\
\hline
$a$ (\AA)  &  6.26  &  7.64  &  7.62 & 10.63  \\
$b$ (\AA)  &  7.84  &  7.70  &  7.76 &  5.50  \\
$c$ (\AA)  & 14.52  & 14.02  & 14.37 & 14.88  \\
$\alpha$ ($^{\circ}$)  & 76.2  &  90.9  &  91.9 &  90.0  \\
$\beta$ ($^{\circ}$)   & 88.0  & 102.8  & 102.0 &  97.5  \\
$\gamma$ ($^{\circ}$)  & 84.7  &  95.2  &  94.8 &  90.0  \\
\end{tabular}
\end{ruledtabular}
\end{table}

\subsection{K$_3$Pentacene compound}

One of the main advantages of our favorite intralayer doping is that it easily allows additional dopants.
For example, three instead of one or two K atoms can be accommodated within the empty regions of the
herringbone structure. Complete structural data are provided by P-1.cif file
of Supplementary Material\cite{cifs}. 
Top panel of Fig. \ref{estructuras-3} shows the arrangement of one of these herringbone layers.
Notice that when dopants enter this way into the herringbone structure, the
spatial crystal symmetry is preserved, that is, a triclinic $P_{-1}$ space group characterizes the
compound as it already happens for pristine pentacene. Nevertheless, very recent
results reported by Nakagawa {\em et al.}\cite{nakagawa2016} for potassium intercalated pentacene
show important structural and electronic changes for the larger 3:1 stoichiometry.
The crystal undergoes a phase transition from the triclinic structure to a new monoclinic one.
Drastically changing the initial crystal structure of a K$_3$Pentacene compound,
we have achieved the structural stabilization of a brand new phase in which former herringbone
structure reassembles completely leaving layers of slipped parallel pentacene molecules.
A complete characterization of the crystalline structure of this monoclinic phase
is given by C2m.cif file\cite{cifs}. 
Besides, bottom panel of Fig. \ref{estructuras-3} allows a visual first impression of the important
structural changes. Layers show a kind of rectangle centered structure which can be considered
typical of an ionic compound. While
lattice parameters of the herringbone phase are not so different from the parameters
of the pristine structure, changes needed to arrive to the monoclinic lattice
are very important (see Table II). Agreement with experimentally determined lattice parameters is
reasonable (see data in Table II of Ref.\onlinecite{nakagawa2016}).
Perhaps the most noticeable change is the perfect parallel arrangement of pentacene molecules.
It allows larger hopping values among pentacene LUMO (LUMO+1) states and, therefore, wider
half-filled bands that should allow metallic electronic transport properties. 
Summarizing, our numeric search provides two quite different crystalline structures for K$_3$Pentacene in
agreement with recent experimental findings\cite{nakagawa2016}.
Although the new monoclinic phase is energetically unfavorable (see data in Table I),
detailed kinetic processes could produce this very regular structure.

Let's discuss relevant differences in the corresponding band structures.
DOS for both of them are shown in Fig. \ref{K3pentaceno}.
Band structure of K$_3$Pentacene in the herringbone structure closely resembles the result
obtained for smaller doping in a similar structure.
Occupation is the main difference. Bands are assigned to particular molecular orbitals
of pentacene in the same way. As it happens for smaller doping,
relevant bands are quite narrow. Therefore, magnetic instabilities
should be expected. Actually, both B$_{\rm 2u}$ and B$_{\rm 1g}$ polarize completely to gain about 0.05 eV
per cell. As it happened at lower doping, we get a ferromagnetic solution due to the initial
polarization conditions. An antiferromagnetic solution showing opposite spin polarization in the
two pentacene molecules forming the unit cell is quite possible although the expected energy
change would be not significant at our computational accuracy level.
Old results by Mori and Ikehata\cite{mori1997} have predicted magnetic phases at low temperatures.
The appearance of spin polarization has also been investigated for the new monoclinic
phase. In this case a ferromagnetic solution is still possible but energetically unlikely.
Table I compiles energies and total cell spin polarization for both structures.
At this point is it possible to predict metallicity for the monoclinic structure.
While the herringbone structure shows a robust spin polarization
giving rise to an insulating DOS ((b) and (b') panels of Fig. \ref{K3pentaceno}),
the smaller value of spin polarization allows a clear metallic Fermi level in the
monoclinic phase. Actually, the broadening of the B$_{\rm 1g}$ band comes
from a not very large $\pi-\pi$ overlap between LUMO+1 states of slipped parallel pentacene molecules.
Although it is considerably larger than for the herringbone structure, it is not
so large as to definitely discard any many-body instability. 

\begin{figure}
\includegraphics[width=0.9\columnwidth]{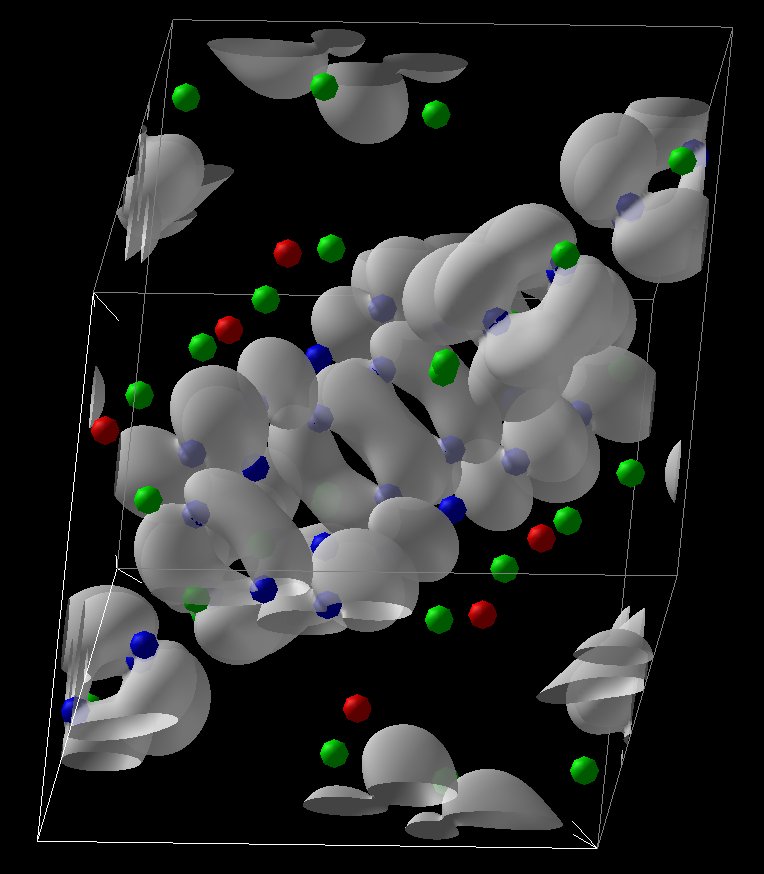}
\caption{(Color online) Iso-surface corresponding to the charge density of
K$_3$Pentacene electrons in bands 129 and 130 (B$_{\rm 1g}$), i.e.
the last two K electrons donated to pentacene
(iso-surface value is 6).
}
\label{parte-ocupada-ultima-banda}
\end{figure}

Densities of charge corresponding to the electrons donated by K have been plotted
for the most stable herringbone structure.
Figs. \ref{127-128} and \ref{parte-ocupada-ultima-banda} 
give isosurfaces than can nicely be interpreted as corresponding to LUMO and LUMO+1 pentacene states.
Therefore our previous assignation of doping bands to particular molecular orbitals
is confirmed. A similar result has been obtained for the monoclinic structure.

An estimation of the powder patterns corresponding to the two structures stabilized in our work can be
easily obtained from the given structural {\em cif} files. Unfortunately, a detailed comparison with
experimentally collected data is not possible without a reinteerpretation of data based in the
new proposed theoretical model.


\subsection{Formation energies}

Formation energies of the K$_x$Pentacene phases studied in our work can be calculated using the data
given in Table I. For example, the formation energy of K$_2$Pentacene from metallic potassium and
pure pentacene is obtained by:

$$
 E({\rm K_2Pentacene}) - E({\rm Pentacene}) - 4 E({\rm K}) ,
$$
giving -5.09 eV
per cell whereas the formation energy of the same compound by the addition of potassium to
KPentacene is given by:

$$
 E({\rm K_2Pentacene}) - E({\rm KPentacene}) - 2 E({\rm K}) ,
$$
resulting in -3.12 eV per cell.
Table III compiles the values corresponding to the K$_x$Pentacene phases preserving
the herringbone structure. All compounds are stable although K$_2$Pentacene shows a preferred
stability as the second column of the Table points out. Actually,
the separation of KPentacene crystal into pure potassium and K$_2$Pentacene phases involves an
energy gain of +0.58 eV per cell. 


\begin{table}
\caption{Formation energies preserving the herringbone structure. Data from
Table I have been used. Energies are given in eV per crystal cell.
The second column gives formation energies relative to
the previous compound in order to emphasize relative stabilities.}
\begin{ruledtabular}
\begin{tabular}{ccc}
KPentacene            &  -1.96 &  -1.96 \\
K$_{2}$Pentacene      &  -5.09 &  -3.12 \\
K$_{3}$Pentacene      &  -5.55 &  -0.46 \\
\end{tabular}
\end{ruledtabular}
\end{table}

\section{Concluding Remarks}

Stable structural phases of K doped pentacene have been theoretically analyzed using a precise
computational scheme. From the point of view of thermodynamic stability, potassium does
always enter into the herringbone structure of the pristine pentacene crystal
in an intralayer manner. Other structural
alternatives like interlayer doping produce higher values of the cell energy.
Extra electrons populate pentacene LUMO firstly and pentacene LUMO+1 afterward. Simultaneously,
they lead to a subtle molecular rearrangement of the herringbone structure
that maximizes symmetry and produces very narrow bands. Therefore, important correlation effects have
to be expected that open the door to either magnetic insulating phases as
shown in our work or more exotic solutions.
On the other hand, kinetically driven reaction pathways may make it possible the stabilization of
a new monoclinic structure
showing about 0.77 eV higher energy per cell. If this happened, our results point to a
metallic band structure for this crystalline phase. In any case, a detailed experimental determination
of the structure of doped pentacene is needed before closing the quest for a metal based on this PAH.

\begin{acknowledgments}
Financial support by the Spanish Ministry of Economy and Competitiveness
(FIS2015-64222-C2-1-P and MAT2016-78625-C2-2-P)
and the University of Alicante is gratefully acknowledged.
\end{acknowledgments}


\begin{thebibliography}{99}
\bibitem{mori-red}
All three lattice constants are larger in the doped sample although
the increase of $c$ is ten times larger than $a$ and $b$ increases.
The discussion of the authors is somewhat ambiguous, doping is
roughly between pentacene layers but some amount goes intralayer.
\bibitem{albert}
These two phases have been referred as pentacene-HT (also called phase C after Campbell's first
report (Ref. \onlinecite{pentaceno-fase-HT}))
and pentacene-LT (or phase-H after Holmes's first report (Ref. \onlinecite{pentaceno-fase-LT}).
A detailed discussion of pentacene polymorphism
is given in Refs. \onlinecite{polymorphism} and \onlinecite{LT-HT}. 
Comparatively, the herringbone structure within the layer is virtually identical for both polymorphs,
while the interlayer spacing, experimentally identified as $d(001)$, is slightly longer in pentacene-HT
($\approx \!\!14.4$ \AA~) respect to the most common pentacene-LT phase ($\approx \!\!14.1$ \AA ~in average).
This experimental distance has been often used to identify each crystallographic phase since their
respective unit cell volumes are very similar.
\bibitem{tumbados}
We have stacked the dimers in a herringbone pattern in which long axis of pentacene
molecules are approximately parallel. Nevertheless, Fig. 9 of Ref.(\onlinecite{pentaceno-dimeros})
shows dimers having long axis that are almost perpendicular. This stacking is quite different from the
original pristine structure. It can be discarded based on energy considerations since our calculations
give a difference of about 0.51 eV per doubled cell between both stacking ways.
\end{thebibliography}
\end{document}